\newcommand{\beq}{\begin{equation}}
\newcommand{\eeq}{\end{equation}}

\newcommand{\id}
 {i\kern.06em\hbox{\raise.25ex\hbox{$/$}\kern-.60em$\partial$}}

\newcommand{\bs}{/\kern-.52em b}
\newcommand{\qs}{/\kern-.52em s}

\newcommand{\dd}
{\kern.06em\hbox{\raise.25ex\hbox{$/$}\kern-.60em$\partial$}}

\newcommand{\ep}{\epsilon}

\documentstyle[12pt]{article}
\begin{document}
\title{Two Theorems on Pseudo-spin in the Hubbard Model
\thanks{On leave of absence from the Physics Department,
Shanghai University, 201800, Shanghai, China}}
\author{{Sze-Shiang Feng,}\\
1.{\small {\it CCAST(World Lab.), P.O. Box 8730, Beijing 100080}}\\
2.{\small {\it Department of Asrtronomy and Applied Physics}}\\
  {\small {\it University of Science and Technology
of China, 230026, Hefei, China}
}\\e-mail:sshfeng@yahoo.com}
\maketitle
\baselineskip 0.3in
\begin{center}
\begin{minipage}{135mm}
\vskip 0.3in
\baselineskip 0.3in
\begin{center}{\bf Abstract}\end{center}
  {An inequality of the eigenvalues of the reduced density matrix
$\rho_2$
  at finite temperature in the Hubbard model is obtained
  by means of the Bogolyubov inequality. The quasi-average
of $\tilde{S}^{+}$ in a simple symmetry-breaking perturbation 
of the Hamiltonian for a bipartite lattice is shown to be zero.
  \\PACS number(s): 75.10.Lp,71.20.Ad,74.65.+n,74.20.Mn
   \\Key words: Hubbard model, pseudo-spin}
\end{minipage}
\end{center}
\vskip 1in
\indent Hubbard models and its extensions such as the SO(5) model\cite{s1}
are currently believed to be able to account for the high-$T_c$
superconductivity of cuprates. The possibility of existence
at finite temperature 
of $s$-wave pairing, generalized $\eta$-pairing and $d_{x^2-y^2}$
in one and two dimensional Hubbard model and Anderson model
seems to have been be ruled out\cite{s2}(of course, the conclusions are
only valid with respect to the special kind of symmetry-breaking perturbations,
there may exist some other kinds of perturbations for which the quasi-average
of the order parametres do not suffer from those limitations, and thus are possibly
non-vanishing).
On the other hand, there do exist some states in some Hubbard
models exhibit superconductivity\cite{s3}\cite{s4}.
Therefore, the relevence to superconductivity of Hubbard models still
calls for further investigations. A convenient concept is off-diagonal
long-range order (ODLRO) discussed in detail in\cite{s5}
which can be studied by $\eta$-pairing \cite{s3}and its variants.
The existence of ODLRO can be recognized if the largest eigenvalue
of the two-body reduced density matrix $\rho_2$ is of the order of the total
number of the fermions\cite{s5}.
A novel property of $\eta$-pairing is that it constutites a pseudo-spin
algebra. Lieb  used this algebra and partial particle-hole transform
and exactly proved for the first time that ferro-magnetism
may exist in itinerant electrons
\cite{s6}. Base on Lieb's results, it is shown that for some specific
lattice structure and electron-filling, the ground-state of negative-$U$
Hubbard may support ODLRO\cite{s4}. In this letter, we first
prove an inequality of the eigenvalues of $\rho_2$ by means of
Bogolyubov inequality.
 Then we show by means of fluctuation-
dissipation theorem that the grand canonical quasi-average of $\tilde{S}^+$
vanishes for both positive and negative-$U$ Hubbard models.  \\
\indent The grand-canonical Hamiltonian for the Hubbard model is
\beq
K=\sum_{(ij)}\sum_{\sigma}t_{ij}c^{\dag}_{i\sigma}c_{j\sigma}+
U\sum_in_{i\uparrow}n_{i\downarrow}-\mu N
\eeq
where $c^{\dag}_{i\sigma}$ and $c_{i\sigma}$ are the creation
and annihlation operators of the electrons with spin $\sigma=
\uparrow, \downarrow$ at site $i$. The hopping matrix
$\{t_{ij}\}$ are required
to be real and symmetric. The number operators are $n_{i\sigma}
=c^{\dag}_{i\sigma}c_{i\sigma}$, while the $U$ denotes the on-site
Coulomb interaction. It is further assumed that the the lattice
$\Lambda$
is bipartite in the sense that it
can be devided into sublattices {\bf A} and {\bf B}, i.e.
$\Lambda={\bf A}\cup{\bf B}$, such that $t_{ij}=o$ whenever
$\{ij\}\in {\bf A}$ or $\{ij\}\in {\bf B}$.
As usual, the spin ${\bf S}$ and pseudo-spin ${\bf \tilde{S}}$ 
 are defined as follows
\beq
S^{+}=\sum_{i\in\Lambda}c^{\dag}_{i\uparrow}c_{i\downarrow},
S^{-}=\sum_{i\in\Lambda}c^{\dag}_{i\downarrow}c_{i\uparrow},
S^z=\frac{1}{2}\sum_{i\in\Lambda}(c^{\dag}_{i\uparrow}
c_{i\uparrow}-c^{\dag}_{i\downarrow}c_{i\downarrow})=\frac{1}{2}
(N_{\uparrow}-N_{\downarrow})
\eeq
\beq
 \tilde{S}^+=\sum_{i\in\Lambda}\epsilon(i)c_{i\uparrow}c_{i\downarrow}
,\,\,\,\,  \tilde{S}^-=\sum_{i\in\Lambda}\epsilon(i)
c^{\dag}_{i\downarrow}
c^{\dag}_{i\uparrow},\,\,\,\,\,
 \tilde{S}^z=\frac{1}{2}\sum_{i\in\Lambda}(1-n_{i\uparrow}
-n_{i\downarrow})
\eeq
where $\epsilon(i)=1 $when $i\in{\bf A}$ and $-1$ when
$i\in{\bf B}$
.Both the spin and the pseudo-spin operators
constitute SU(2) algebra and they
commute with each other, i.e. $[{\bf \tilde{S}}, {\bf S}]=0$ ,
so they form an SU(2)$\otimes$ SU(2) algebra.
It is not difficult to show that  $[H, {\bf \tilde{S}}^2]=
[H, {\bf S}]=[H,\tilde{S}_z]=0$. Therefore, Hubbard model
 enjoyes $SU(2)\otimes U(1)\otimes U(1)
$ symmetry.  Yang and Zhang\cite{s7} showed that
ODLRO exists whenever the expectation value of
${\bf \tilde{S}}^2-
\tilde{S}_z^2$ is of order $N_{\Lambda}^2$, where
$N_{\Lambda}$
is the number of the sites of the lattice considered.\\
\indent The well-known Bogolyubov inequality\cite{s8} says that for any two
operators $C$ and $Q$, the following inequality always holds
\beq
<CC^{\dag}+C^{\dag}C>\ge\frac{2}{\beta}\frac{|<[Q(t), C(t)]>|^2}{|
<[Q(t), [Q^{\dag}(t),K]]>|^2}
\eeq
where $<{\cal O}>$ denote the statistical average of the mechanical quantity
${\cal O}$.This inequality has been utilized in\cite{s2}.
 We employ it here to prove
a relation the eigenvalues of the reduced density $\rho_2$ should satisfy.
Let us recognize some fundamental properties of the eigenvalues before
presenting the relation. Firstly,  it is not difficult to show that
the  onset of ODLRO can be guaranteed if the largest
eigenvalue of either
 $<\ep(j)\eta_j^{\dag}\ep(i)\eta_i>$ or $<\eta_j^{\dag}\eta_i>$, where the
$\eta$ pairing is defined as $\eta_i=c_{i\uparrow}c_{i\downarrow}$,
 is of the order of $N_{\Lambda}$. In fact, we have the following lemma.\\
{\bf Lemma 1} {\it The matrices 
 $<\ep(j)\eta_j^{\dag}\ep(i)\eta_i>$ and $<\eta_j^{\dag}\eta_i>$.
have the same eigenvalues.}\\
{\it Proof} Suppose $\sum_i<\eta_j^{\dag}\eta_i>u_i=\lambda u_j,$ then
we have
\beq
\sum_i<\ep(j)\eta_j^{\dag}\eta_i\ep(i)>\ep(i)u_i=\ep(j)\sum_i<\eta^{\dag}_j
\eta_i>u_i=\lambda\ep(j)u_j
\eeq
and this proves the lemma. \hfill{Q.E.D.}\\
\indent Secondly, the eigenvalues are non-negative.\\
{\bf Lemma 2} {\it The matrix  $<\ep(j)\eta_j^{\dag}\ep(i)\eta_i>$ is positive
semi-definite.}\\
{\it Proof} For an arbitrary vector $u_i$, define $O=\sum_i\ep_i\eta_iu_i$,
 then
\beq
\sum_{i,j}u^*_j<\eta^{\dag}_j\ep(j)\ep(i)\eta_i>u_i=<O^{\dag}O>\ge 0
\eeq
Therefore, all the eigenvalues are non-negative.\hfill{Q.E.D.}\\
\indent Beside the above two lemmas, we will use the next lemma in the
following.\\
{\bf Lemma 3} {\it Let $A$ be an $N\times N$ matrix whose entries
satisfy $a_{mn}=a_{m-n}$. Then all the eigenvalues of $A$ are given
by $\lambda_q=\frac{1}{N}\sum^N_{m=1}\sum^N_{n=1}a_{mn}e^{iq(m-n)}$, where
$q=\frac{2\pi}{N}k$ with $k=0,1,\cdots, N-1$. }\\
This lemma has been used in \cite{s9} and \cite{s10} and a proof is presented
in \cite{s10}. Now we are ready to present our first result\\
{\bf Theorem 1} {\it The eigenvalues of$<\eta^{\dag}_j\eta_i>$ can be denoted
by wave vectors ${\bf k}$ as $\lambda_{\bf k}$, ${\bf k}$
 belong to the first Brillouin zone. If the system is translationally
invariant and the lattice is bipartite, the following inequality holds
 \beq
 \lambda_{\bf k}\ge|x|\frac{1}{\beta}\frac{1}{|2\mu-U|}
\delta_{{\bf k},0}-\frac{x}{2}
 \eeq
where $x=1-\rho_e$}.\\
{\it Proof}  For our purpose, we choose
\beq
Q=\tilde{S}^-
\eeq
and
\beq
C_{\bf k}=\sum_je^{i{\bf k}\cdot{\bf R}_j}\ep(j)\eta_j
=\sum_je^{i{\bf k}\cdot{\bf R}_j}\ep(j)c_{j\uparrow}c_{j\downarrow}
\eeq
Making use of the relation $[\eta_i^{\dag}, \eta_j]=\delta_{ij}
(n_{i\uparrow}+n_{i\downarrow}-1)
$, we have
\beq
[Q, C_{\bf k}]=\sum_je^{i{\bf k}\cdot{\bf R}_j}(n_j-1)
\eeq
Since
\beq
[\tilde{S}^-, K]=(2\mu-U)\tilde{S}^-
\eeq
we have
\beq
[Q^{\dag}, K]=(U-2\mu)\tilde{S}^+
\eeq
\beq
[Q,[Q^{\dag},K]]=2(2\mu-U)\tilde{S}_z
\eeq
By direct calculation, we have
\beq
C^{\dag}_{\bf k}C_{\bf k}+C_{\bf k}C^{\dag}_{\bf k}
=2\sum_{ij}e^{i{\bf k}\cdot({\bf R}_i-{\bf R}_j)}
\ep(i)\ep(j)\eta^{\dag}_{j}\eta_i-\sum_i(n_i-1)
\eeq
Therefore, using Bogolyubov inequality, we have
$$
2\sum_{ij}e^{i{\bf k}\cdot({\bf R}_i-{\bf R}_j)}
\ep(i)\ep(j)<\eta^{\dag}_j\eta_i>-\sum_i<n_i-1>$$
\beq
\ge
\frac{2}{\beta}\frac{|<\sum_je^{i{\bf k}\cdot{\bf R}_j}(n_j-1)>|^2}
{|2(2\mu-U)||<\tilde{S}_z>|}
=\frac{2}{\beta}\frac{\sum_{ij}e^{i{\bf k}\cdot({\bf R}_i-{\bf R}_j)}
<n_i-1><n_j-1>}{2|2\mu-U|\cdot \frac{1}{2}\sum_j|<1-n_j>|}
\eeq
Since we consider the translationally invariant system,
$<n_i>=N_e/N_{\Lambda}=\rho_e$, accordingly
\beq
2\sum_{ij}e^{i{\bf k}\cdot({\bf R}_i-{\bf R}_j)}
\ep(i)\ep(j)<\eta^{\dag}_j\eta_i>+N_{\Lambda}x
\ge
\frac{2}{\beta}\frac{\sum_{ij}e^{i{\bf k}\cdot({\bf R}_i-{\bf R}_j)}
x^2}{|2\mu-U|\cdot N_{\Lambda}|x|}
=\frac{2}{\beta}|x|\frac{\delta_{{\bf k},0}N_{\Lambda}}{|2\mu-U|}
\eeq
where we have used that $\sum_{ij}e^{i{\bf k}\cdot({\bf R}_i-{\bf R}_j)}
=N_{\Lambda}^2\delta_{{\bf k},0}$. Or
\beq
2\sum_{ij}e^{i{\bf k}\cdot({\bf R}_i-{\bf R}_j)}
<\ep(j)\eta^{\dag}_j\ep(i)\eta_i>
\ge\frac{2}{\beta}|x|\frac{\delta_{{\bf k},0}N_{\Lambda}}{|2\mu-U|}-N_{\Lambda}x
\eeq
On the other hand, lemma 3 tells us that
the eigenvalues of the $N_{\Lambda}\times N_{\Lambda}$
matrix $<\ep(j)\eta^{\dag}_j\ep(i)\eta_i>$ are
\beq
\lambda_{\bf k}=\frac{1}{N_{\Lambda}}\sum_{ij}e^{i{\bf k}\cdot({\bf R}_i-{\bf R}_j)}
<\ep(j)\eta^{\dag}_j\ep(i)\eta_i>
\eeq
and this proves the theorem.\hfill{Q.E.D.}\\
\indent We make some remarks here. In the case of at half-filling, $x=0$.
in this case, the theorem provides no new information because
lemma 2 has told us that all the eigenvalues are non-negative.
It should be mentioned that Tian showed in\cite{s10} that at half-filling,
for the negative$-U$ case, the lowest eigenvalue is zero in the ground-state.
Our result is in accordance with Tian's, though the zero-temperature limit
of a grand-canonical average does not necessarily coincide with the ground
state expectation value in general such as\cite{s11} and \cite{s4}.
But, if the system is at over half-filling, $x$ will be negative and
the $r.h.s.$ of (7) is positive. in this case, the theorem tells us
that there exists lower positive bound to each eigenvalue. Unfortunately,
the lower bound to the largest eigenvalue the theorem provides
is only a constant other than a number proportional to $N_e$.\\
\indent Our second result is\\
{\it {\bf Theorem 2} 
Assuming the lattice is bipartite, for the U(1) symmetry-breaking
Hamiltonian }
\beq
H=\sum_{(ij)}\sum_{\sigma}t_{ij}c^{\dag}_{i\sigma}c_{j\sigma}+
U\sum_in_{i\uparrow}n_{i\downarrow}+\Delta\tilde{S}^++\Delta^*\tilde{S}^-
\eeq
{\it we have for the grand canonical average of } $\tilde{S}^+$
\beq
\lim_{\Delta\rightarrow 0}\lim_{N_{\Lambda}\rightarrow\infty}\frac{1}{N_{\Lambda}}
<\tilde{S}^+>=0
\eeq
{\it Proof} The grand canonical Hamiltonian
is
$K=H-\mu{\hat N}$. 
We use the fluctuation-dissipation theorem as in \cite{s11}.
to calculate
$<\tilde{S}^+\tilde{S}^->$. 
The evolution
equation
of  the double-time Green function
$\ll{\bf \tilde{S}}^-|{\bf \tilde{S}}^+\gg_{\omega}$. 
is\cite{s12}
\beq
\omega\ll{\bf \tilde{S}}^-|{\bf \tilde{S}}^+\gg_{\omega}
=<[{\bf \tilde{S}}^-, {\bf \tilde{S}}^+]>+\ll[{\bf \tilde{S}}^-,
K]|{\bf \tilde{S}}^+\gg_{\omega}
\eeq
It can be calculated directly that
\beq
[{\bf \tilde{S}}^-,K]=(2\mu-U){\bf \tilde{S}}^--2\Delta\tilde{S}_z
\eeq
Accordingly, we have
\beq
\omega\ll\tilde{S}^-|\tilde{S}^+\gg_{\omega}=-2<\tilde{S}_z>
+(2\mu-U)\ll\tilde{S}^-|\tilde{S}^+\gg-2\Delta\ll\tilde{S}_z|
\tilde{S}^+\gg_{\omega}
\eeq
Using again the equation of motion, we can obtain
\beq
\omega\ll\tilde{S}^-|\tilde{S}^+\gg_{\omega}
=-<\tilde{S}^+>-\Delta\ll\tilde{S}^+|\tilde{S}^+\gg_{\omega}
+\Delta^*\ll\tilde{S}^-|\tilde{S}^+\gg_{\omega}
\eeq
and
\beq
\omega\ll\tilde{S}^+|\tilde{S}^+\gg_{\omega}
=-(2\mu-U)\ll\tilde{S}^+|\tilde{S}^+\gg_{\omega}
+2\Delta^*\ll\tilde{S}^z|\tilde{S}^+\gg_{\omega}
\eeq
These three equations eq.(23)-eq.(25) are enough to determine the three
unknowns
$\ll\tilde{S}^-|\tilde{S}^+\gg_{\omega},
\ll\tilde{S}^+|\tilde{S}^+\gg_{\omega},
$ and $\ll\tilde{S}^z|\tilde{S}^+\gg_{\omega}$. Especially, we
have
\beq
\ll\tilde{S}^-|\tilde{S}^+\gg_{\omega}=\frac{c_1}{\omega}
+\frac{c_2}{\omega-\omega_2}+\frac{c_3}{\omega-\omega_3}
\eeq
where
\beq
c_1=[2\Delta(2\mu-U)<\tilde{S}^+>-4|\Delta|^2<\tilde{S}_z>]
\frac{1}{4|\Delta|^2-(2\mu-U)^2}
\eeq
\beq
c_2=\frac{1}{\omega_2(\omega_2-\omega_3)}\{2\Delta<\tilde{S}^+>
(\omega_2+2\mu-U)-2[\omega_2(\omega_2+2\mu-U)+2|\Delta|^2]<\tilde{S}_z>\}
\eeq
\beq
c_3=\frac{1}{\omega_3(\omega_3-\omega_2)}\{2\Delta<\tilde{S}^+>
(\omega_3+2\mu-U)-2[\omega_3(\omega_3+2\mu-U)+2|\Delta|^2]<\tilde{S}_z>\}
\eeq
\beq
\omega_{2,3}=\pm\sqrt{(2\mu-U)^2-4|\Delta|^2}
\eeq
Thus from the fluctuation-dissipation theorem
\beq
<\tilde{S}^+\tilde{S}^->=\frac{i}{2\pi}\int^{+\infty}_{-\infty}d\omega
\frac{\ll\tilde{S}^-|\tilde{S}^+\gg_{\omega+i\eta}-
\ll\tilde{S}^-|\tilde{S}^+\gg_{\omega-i\eta}}{e^{\beta\hbar\omega}-1}
\eeq
we know that $c_1$ must vanish, otherwise,
there will be $<\tilde{S}^+\tilde{S}^->=\infty$
for finite $N_{\Lambda}$. Thus
\beq
\frac{1}{N_{\Lambda}}<\tilde{S}^+>=\frac{2\Delta^*}{2\mu-U}\frac{1}{N_{\Lambda}}
<\tilde{S}_z>
\eeq 
Taking the limit in (20), one can immediately arrive at the conclusion
\hfill{ Q.E.D.}\\.
\indent According to Bogolyubov\cite{s13}, symmetry-breaking
properties should be discussed by quasi-averages, i.e. giving some
symmetry-breaking perturbation and letting it  
vanish finally. Theorem 2 states that the quasi-average of
$\tilde{S}^+$ is zero. This does not mean the  pairing
$\lim_{\Delta\rightarrow 
0}\lim_{N_{\Lambda}\rightarrow\infty}<c_{i\uparrow}c_{i\downarrow}>=0$.
If the two sub-lattices are homogeneous, denote
$\tilde{s}^+_{A/B}=<c_{i\uparrow}c_{i\downarrow}> $ for
 $i\in {\bf A}/{\bf B}$. Theorem 2 implies that
$\rho_A\tilde{s}^+_A-\rho_B\tilde{s}^+_B=0$, where $\rho_{A/B}=
N_{A/B}/N_{\Lambda}$.
\indent We make some conclusional discussions. It can be understood
that we  can not obtain a
lower limit of the order of $N_{\Lambda}$ for the largest eigenvalue
of the reduce density $\rho_2$  by means of the Bogolyubov 
identity (4) because the times of sum $\sum_i$
in both the numerator
and the denominator in the right hand side of (4) are equal 
. Thus the value will not be of order $N_{\Lambda}$.
As the conclusions in \cite{s2} hold only
for the specific quasi-averages therein,
our theorem 2 refers to only a special way of symmetry breaking, or
a special quasi-average too. It can not surely rule out the possibility
for non-vanishing $<\tilde{S}^{+}>$ for other quasi-averages.
Similar discussions seem to apply to other models such as
Kondo lattice model and periodic Anderson model. Finally,
we would like to stress that both theorems do not rule out the possible
existence of ODLRO in Hubbard models.

\vskip 0.3in
\underline{Acknowledgement}  The author is grateful to Prof. F. Mancini
for reading the manuscript and helpful discussions and suggestions.
This work was supported by the Funds for Young Teachers of Shanghai Education
Committee and in part by the National Science Foundation under
Grant. 19805004.\\

\end{document}